\documentclass[twocolumn,showpacs,aps,floatfix,superscriptaddress,amsmath,amssymb,amsfonts,longbibliography]{revtex4-1}
\usepackage{times}
\usepackage[varg]{txfonts}
\usepackage{textcomp}
\usepackage{graphicx}
\usepackage{epstopdf}
\usepackage{subfigure}
\usepackage{tabu}
\usepackage{color}
\usepackage[colorlinks=true,citecolor=blue,urlcolor=blue,linkcolor=blue,breaklinks,hyperindex]{hyperref}
\usepackage{braket}
\usepackage{overpic}
\usepackage{amssymb}
\usepackage{appendix}
\usepackage{epsfig}

\allowdisplaybreaks

\begin{document}

\title{Spin Glass in the Bond-Diluted $J_{1}-J_{2}$ Ising Model on the Square Lattice}
\author{Yining Xu}
\affiliation{State Key Laboratory of Optoelectronic Material and Technologies, School of Physics, Sun Yat-sen University, Guangzhou 510275, China}
	
\author{Dao-Xin Yao}
\email[Corresponding author:]{yaodaox@mail.sysu.edu.cn}
\affiliation{State Key Laboratory of Optoelectronic Material and Technologies, School of Physics, Sun Yat-sen University, Guangzhou 510275, China}

\date{\today}

\begin{abstract}

We use Monte Carlo (MC) methods to simulate a two-dimensional (2D) bond-diluted Ising model on the square lattice which has frustration between the nearest-neighbor interaction $J_{1}$ and the next-nearest-neighbor interaction $J_{2}$. In this paper, we use the parallel tempering algorithm to study the thermodynamics for different diluted ratio $x$ and give the phase diagram. The presence of both frustration and disorder results in a spin-glass phase, which exists between the stripe antiferromagnetic phase and the N\'{e}el phase. We present the ground-state energy of $T \rightarrow 0$ and the size-dependence of Edwards-Anderson (EA) order parameter for the spin glass phase. By scaling the mean energy and the EA order parameter from the simulated annealing with the Kibble-Zurek (KZ) mechanism, we obtain two different dynamic exponents $z_{E}$ and $z_{q}$ for the spin glass phase. Experimentally, this model has close implication with the $\mathrm{FeAs}$ plane of the iron-based superconductor $\mathrm{BaFe}_{2}(\mathrm{As}_{1-x}\mathrm{P}_{x})_{2}$, where a spin-glass like phase was found.

\end{abstract}

\maketitle

\section{Introduction}

Lots of theoretical and experimental efforts have been dedicated to study the properties of spin glass~\cite{Binder1986SG}, in which spins are freezing and disordered. The theoretical model of spin glass was proposed by Edwards and Anderson~\cite{Edwards1975SG}. The mean-field theory is the original idea for studying spin glass, models like Edwards-Anderson model and Sherrington-Kirkpatrick model~\cite{SKmodel1975} are based on it. Until the Almeida-Thouless line~\cite{ATline1978} is found, the replica symmetry breaking begin to be considered in spin glass. In experiment, some of the characteristic phenomena have been observed in spin glass, such as the rather sharp cusp in the frequency-dependent susceptibility in low fields~\cite{Cannella1972Magnetic} and remanence~\cite{THOLENCE1974SUSCEPTIBILITY,Ocio1985Observation} and hysteresis below the freezing temperature~\cite{Monod1979Magnetic,Prejean1980Hysteresis}. The behaviors of spin glass can be observed in experiment by methods such as nuclear magnetic resonance (NMR) and neutron scattering (NS). Among the spin-glass systems, the two-dimensional (2D) Ising spin glass (ISG) is a special one since it only exists at temperature $T = 0$~\cite{glass1977Villain}. The commonly discussed 2D ISG models, for example, square lattice with Gaussian or bimodal couplings, which contain randomly distributed ferromagnetic and antiferromagnetic interactions. In recent years, there are many studies~\cite{excitations2012,energyentropy2012,glassy2010,Lowtemperature2004,XuNa2017Dual} focusing on the low-temperature behaviors and phase transition on 2D ISG .

Besides the frustrated interactions, disorder plays important role in the spin glass such as dilutions. Bond dilution can be realized by changing the interactions between two spins. Site dilution can be achieved by removing or changing a certain portion of the spins on the lattice. A considerable number of investigations~\cite{bonddilute1978,3DdilutedIsing1999,sitediluted2008,sitediluted2010} focusing on locating transition point and critical behaviors by renormalization-group methods and Monte Carlo methods. Diluted spin models can be realized in many materials. For example, $\mathrm{Fe}_{x}\mathrm{Zn}_{1-x}\mathrm{F}_{2}$ and $\mathrm{Mn}_{x}\mathrm{Zn}_{1-x}\mathrm{F}_{2}$ which are prepared by a substitution of the nonmagnetic isomorph $\mathrm{ZnF}_{2}$ for the magnetic ones $\mathrm{FeF}_{2}$($\mathrm{MnF}_{2}$), can be described as a three-dimensional diluted Ising model~\cite{diluted2003Folk}. Modulation of pairing symmetry with bond dilution in iron-based superconductors has been studied by Ref.~\cite{Kang2017superconductors}. In this paper, we study a 2D bond-diluted Ising model with the nearest-neighbor interaction $J_{1}$ and the next-nearest-neighbor interaction $J_{2}$ on the square lattice, which is similar to the $\mathrm{FeAs}$ plane of the iron-based superconductor $\mathrm{BaFe}_{2}(\mathrm{As}_{1-x}\mathrm{P}_{x})_{2}$. Since the spin size of the $\mathrm{Fe}$ atoms are generally large, we can use Ising spins to describe the magnetism. The random distribution of $\mathrm{P}$ atoms can lead to an effective $J_{2}$ dilution on the square lattice.  A spin-glass-like behavior in $\mathrm{BaFe}_{2}(\mathrm{As}_{1-x}\mathrm{P}_{x})_{2}$ was found by nuclear magnetic resonance (NMR) and triple-axis spectrometer (TRISP) measurements \cite{MagneticBaFeAsPH2015}, where the superconductivity also happens \cite{BaFeAsP2009,Unconventional2010Nakai}. However a systematic study on the magnetism was lack.

Fundamentally, it is also interesting to understand the thermodynamics of the 2D bond-diluted $J_{1}-J_{2}$ Ising model, which is different from the clean $J_{1}-J_{2}$ Ising model~\cite{Oitmaa1981Ising,Jin2013Ising,Guerrero2015Nematic}. Our study shows that the system really has a spin-glass phase which can be controlled by the bond dilution and frustration.

In this paper, we use the highly efficient Monte Carlo (MC) methods (parallel tempering and simulated annealing) to study the 2D bond-diluted $J_{1}-J_{2}$ Ising model. We study the thermodynamics of the ordered phase as the dilution ratio $x$. A spin-glass phase is found in between the stripe antiferromagnetic phase and the N\'{e}el phase. In the spin-glass phase, we find two different dynamic exponents which are obtained from the simulated annealing results of the mean energy and Edwards-Anderson (EA) order parameter. This unusual behavior is similar to the results of the 2D $\pm J$ ISG model in Ref.\cite{XuNa2017Dual}. The phase diagram can help to understand the experimental phase diagram of $\mathrm{BaFe}_{2}(\mathrm{As}_{1-x}\mathrm{P}_{x})_{2}$.

The rest of the paper is organized as follows. In Sec.~\ref{mod-meth} we introduce the model and methods. Numerical results are presented in Sec.~\ref{Nures}, including the spin-glass phase and its dynamical properties. In Sec.~\ref{dscus} we discuss the comparability of our results with the experimental results. Conclusions are given in Sec.~\ref{conclu}. Additional results, which discusses the connections with the 2D $\pm J$ ISG model are given in appendix ~\ref{ap-A}.

\section{Model and methods\label{mod-meth}}

\subsection{Model\label{md}}

\begin{figure}
\centering
{\resizebox*{0.3\textwidth}{!}{\includegraphics{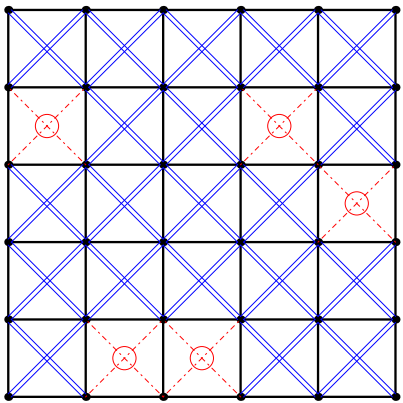}}}
\caption{(Color online). 2D square-lattice Ising model with two types of interaction. The solid thick line represents the nearest-neighbor interaction $J_{1}$. The blue double line represents the next-nearest-neighbor interaction $J_{2}$. The $J_{2}$ always come in pairs in a plaquette, and can be broken in pairs as show by the red dashed line.
\label{mod}}
\end{figure}

We here study a bond-diluted $J_{1}-J_{2}$ Ising model on the 2D square lattice, as shown in Fig.~\ref{mod}. The structure of this model is very similar to the $\mathrm{FeAs}$ plane of the iron-based superconductor $\mathrm{BaFe}_{2}(\mathrm{As}_{1-x}\mathrm{P}_{x})_{2}$. The Hamiltonian of the model is
\begin{equation}
H=J_{1}\sum_{<i,j>} \sigma_{i} \sigma_{j}+J_{2}\sum_{<i',j'>} \delta_{i'j'} \sigma_{i'} \sigma_{j'}, \quad \sigma_{i} = \pm 1,
\end{equation}
where $J_{1}$ is the nearest-neighbor interaction, $J_{2}$ is the next-nearest-neighbor interaction, and $\delta_{i'j'}$ represents the bond dilution ($1$ represents the existence of $J_{2}$-bond, $0$ means the bond dilution, shown in Fig.~\ref{mod}). We define the bond dilution ratio as $x=1-N(J_{2})_{\mathrm{pair}}/N$, where $N$ is the total number of $J_{2}$-bonds without the bond dilution. Here we set $J_{1}=J_{2}=1$,

\begin{figure}
\centering
{\resizebox*{0.3\textwidth}{!}{\includegraphics{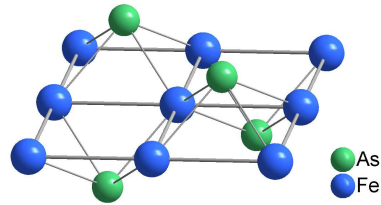}}}
\caption{(Color online). This is the $\mathrm{FeAs}$ plane of $\mathrm{BaFe}_{2}(\mathrm{As}_{1-x}\mathrm{P}_{x})_{2}$. The $\mathrm{Fe}$ atoms form a square lattice. Here the $x$ is the doping ratio of $\mathrm{BaFe}_{2}(\mathrm{As}_{1-x}\mathrm{P}_{x})_{2}$. When $x > 0$, a part of the $\mathrm{As}$ atoms are substituted by the $\mathrm{P}$ atoms.
\label{mod-ex}}
\end{figure}

In $\mathrm{BaFe}_{2}(\mathrm{As}_{1-x}\mathrm{P}_{x})_{2}$, both magnetism and superconductivity occur in the $\mathrm{FeAs}$ plane. In the $\mathrm{FeAs}$ plane, the $\mathrm{As}$ atoms sit alternatively above and below the center of each plaquette on the square lattice which is formed by the $\mathrm{Fe}$ atoms, as shown in Fig.~\ref{mod-ex}. In the $\mathrm{FeAs}$ plane, the $\mathrm{P}$ atoms can be randomly substituted for the $\mathrm{As}$ atoms.

In our study, we consider the spin magnetism on the $\mathrm{Fe}$ atoms. The spin on the $\mathrm{Fe}$ atoms can be considered as Ising spins because of the high anisotropy. The nearest-neighbor interaction between the $\mathrm{Fe}$ atom is defined as the $J_{1}$. The $\mathrm{As}$ atoms generate the superexchange interactions between the $\mathrm{Fe}$ atoms, while the $\mathrm{P}$ atoms can not. An $\mathrm{As}$ atom constructs a pair of $J_{2}$ between four $\mathrm{Fe}$ atoms. When the $\mathrm{As}$ atom is substituted for the $\mathrm{P}$ atom, the $J_{2}$ are broken in pairs. The $x$ is the doping ratio of the $\mathrm{P}$ atoms, which equals to the diluted radio we defined.

In this model, when the $J_{1}$ and $J_{2}$ exist simultaneously, occurring a competitive relationship because the $J_{1}$ and $J_{2}$ both are antiferromagnetic and form a triangular structure. Our random dilution results in a disordered distribution of the $J_{2}$, just as the disordered doping in $\mathrm{BaFe}_{2}(\mathrm{As}_{1-x}\mathrm{P}_{x})_{2}$ material. The competitive interactions cause frustration, and the disorder is introduced into the system by randomly diluting. Due to the combination of frustration and disorder, our study obtain an interesting discovery.

\subsection{Parallel tempering\label{pt}}

For complex systems, the energy landscape has many separated local minima. The simulation of complex systems by using the conventional MC usually requires long relaxation time. In conventional MC studies, simulations of higher temperatures are generally sampled in large volumes of phase space, whereas the low temperature ones maybe trapped in local energy minima during the timescale of a typical computer simulation. To overcome this problem, many different MC methods have been discussed\cite{Hukushima1996Exchange}. The form of parallel tempering MC, now frequently used, can dates back to the study of Geyer\cite{Geyer1991Markov}. In the developmental process of the parallel tempering have many similar forms, such as Replica Monte Carlo\cite{Swendsen1986Replica}, simulated tempering\cite{Enzo1992SIMULATED}, and expanded ensemble method\cite{Lyubartsev1992New}. All of these methods simulate the complex systems over a wide temperature range, helping complex systems to escape from metastable states and speeding up the equilibrium process.

The parallel tempering method we used here allows the system to exchange the complete configuration between different temperatures, ensuring that the lower temperature system can access a set of representative regions of phase space. We briefly summarize the sampling procedure of the parallel tempering method. The operation is implemented in two stages, including simple single-temperature MC stage and parallel tempering stage. In the simple MC stage, $N$ non-interacting replicas of the system are simulated simultaneously by respectively performing single-temperature Metropolis update at $N$ different temperatures, {$T_{1}$,$T_{2}$,...,$T_{N}$}. The parallel tempering stage carries out replica exchange, where two replicas at neighboring temperatures swap the complete configuration. The swapping probability $p_{\mathrm{swap}}$ between two neighboring temperatures, $T_{i}$ and $T_{i+1}$, is given by:
  \begin{equation}
  p_{\mathrm{swap}}=\mathrm{min} \{ 1,\mathrm{exp}[(\frac{1}{T_{i+1}}-\frac{1}{T_{i}})(E_{i+1}-E_{i})] \},
  \end{equation}
where $E_{i}$ and $E_{i+1}$ are the energy of the replica at temperature $T_{i}$ and $T_{i+1}$ respectively.

Now, the parallel tempering MC is widely considered as a powerful method to study the complex systems. In this paper, we try to use the improved form, which adjusts the distribution of the temperatures or the steps of the two stages~\cite{earl2005parallel,katzgraber2006feedback,liu2016role}. These improvements can make the computation more efficient.

\subsection{Simulated annealing \label{an}}

Simulated annealing is so named due to the fact that it has the similar process with physical annealing~\cite{Nikolaev2010Simulated}. In physical annealing, a crystal is heated and then cooled slowly until attaining one of the most common crystal lattice structures, so that the defects of crystal can be removed. If the cooling is sufficiently slow, the final configuration can approach a superior structure. Numerically, simulated annealing establishes the connection between the thermodynamic behaviors of physical annealing and the search for global minima of a discrete optimization problem~\cite{Kirkpatrick1983Optimization,INGBER1993Simulated}.

For the spin systems with rough energy landscape, simulated annealing is a sequential MC process. In the beginning, finding the equilibrium state in initial temperature by using standard Metropolis update is necessity. Then slowly decreasing temperature step by step until the critical temperature, while updating the system in the temperature of every step. Simulated annealing is a powerful algorithm in exploring the energy landscape of complex systems, and capable of escaping from local minimums. Even though the simulated annealing and the parallel tempering play a similar role~\cite{Wang2015Comparing} in detecting the ground states of complex systems, however the differences between them are distinct. Simulated annealing is a non-equilibrium process, which does not give any meaningful results during the annealing process except for the non-equilibrium results obtained from the critical temperature. The non-equilibrium results obtained after the annealing process are related to the annealing velocity and the system size. The study of phase transitions with simulated annealing is based on the Kibble-Zurek mechanism~\cite{Kibble1976,Zurek1985}, which is originally used in the non-equilibrium scaling of the defect density in condensed-matter physics and now is successfully used to describe non-equilibrium physics at both classical and quantum phase transitions.

\section{numerical results\label{Nures}}

\subsection{The ordered phase\label{ord}}
\begin{figure}
\centering
\includegraphics[width=0.43\textwidth,height=5.2in]{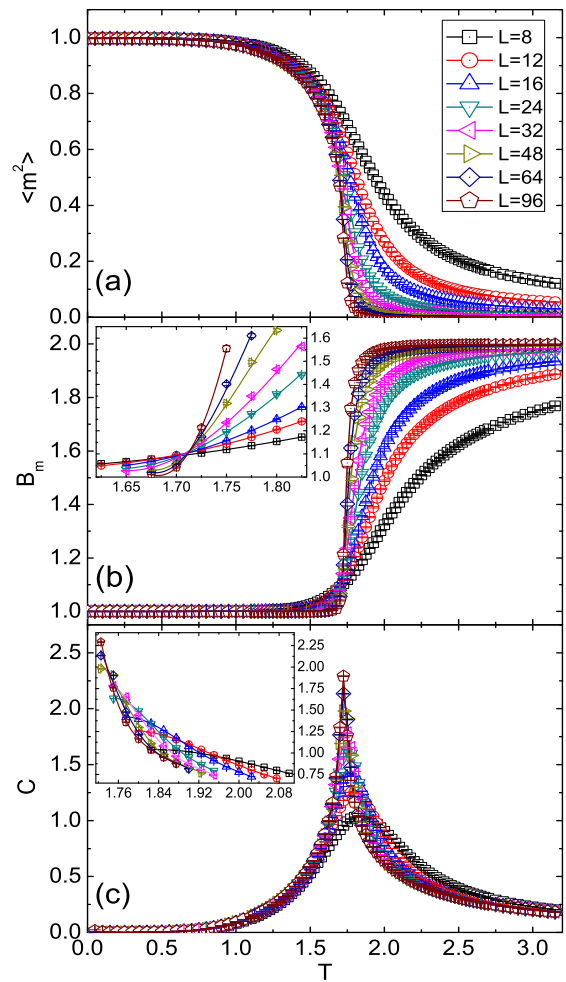}
\caption{(Color online) The results of the $x = 0.1$, where the dilution of the system is not enough to changes the order of the ground state. (a) The order parameter shows that the ground state has a stripe antiferromagnetic order. The results of Binder cumulant are shown in (b), and (c) corresponds to the specific heat. By the polynomial fitting to the data, we obtain the cross-points which corresponds to the critical temperature $T_{c}(L)$. }
\label{order1}
\end{figure}
From the Hamiltonian, we can easily find the ground state of the system is the N\'{e}el state when the dilution ratio $x = 1$, which implies that the system has no $J_{2}$ interaction. When the dilution ratio $x = 0$, the antiferromagnetic interactions $J_{1}$ and $J_{2}$ exist simultaneously, thus introducing frustrations into the system. However the $J_{1}$- and $J_{2}$-bonds are distributed in an ordered pattern, so the ground state of the system exhibits the stripe antiferromagnetic order. In this section, we focus on how the ground state changes as a function of $x$. Under different dilutions, we investigate the critical temperature of the ordered phases to the paramagnetic phase. We use an order parameter $m_{s}$~\cite{Jin2013Ising} to describe the stripe antiferromagnetic order, which can be defined as
  \begin{equation}
  \begin{aligned}
  m_{s}^{2}=&m_{x}^{2}+m_{y}^{2},\\
  m_{x}^{2}=&\frac{1}{N}\sum_{i=1}^N \sigma_{i}(-1)^{x_{i}},\\
  m_{y}^{2}=&\frac{1}{N}\sum_{i=1}^N \sigma_{i}(-1)^{y_{i}}.
  \end{aligned}
  \end{equation}
The N\'{e}el order parameter is defined as
  \begin{equation}
  m_{N}=\frac{1}{N}\sum_{i=1}^N \sigma_{i}(-1)^{x_{i}+y_{i}}.
  \end{equation}

To locate the critical temperatures, we use the Binder cumulant~\cite{vollmayr1993finite} defined as
  \begin{equation}
  B_{m}=\frac{\langle m^{4}\rangle}{\langle m^{2}\rangle^{2}},
  \end{equation}
where $m$ represent $m_{s}$ or $m_{N}$. Specific heat ($C$), is also computed in order to study the phase transition from an ordered state to the paramagnetic state. We show results of $x = 0.1$ in Fig.~\ref{order1} and perform the same analysis for the rest of $x$.

\begin{figure}
\centering
\includegraphics[width=0.45\textwidth,height=1.9in]{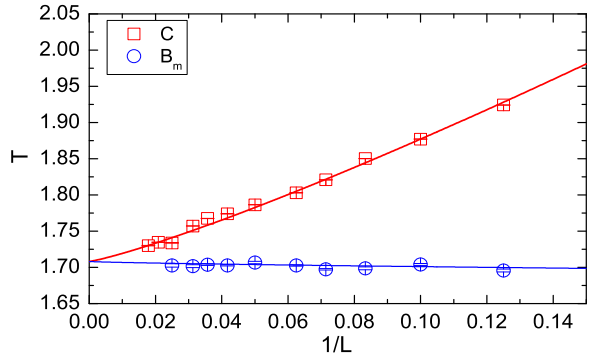}
\caption{(Color online) The $T_{c}(L)$ obtained from the specific heat ($C$) and the Binder cumulant ($B_{m}$). When $x = 0.1$, the system exists the long-range order. Using the form $T_{c}(L)=T_{c}(\infty)+a/L^{b}$, we can extrapolate the $T_{c}(L)$ to infinite size and finally get the corresponding critical temperature $T_{c}(\infty)$ of $x$. Here $a$, $b$ are fitting parameters, and the critical temperature $T_{c}(\infty)=1.708(9)$.}
\label{order2}
\end{figure}

We obtain the critical temperatures $T_{c}(L)$ from the cross-points of the curves for $L$ and $2L$. By performing the power-law fitting, and we obtain the critical temperature $T_{c}$ for the thermodynamic limit. In Fig.~\ref{order2}, we give an instance with $x = 0.1$.

\begin{figure}
\centering
\includegraphics[width=0.5\textwidth,height=3.9in]{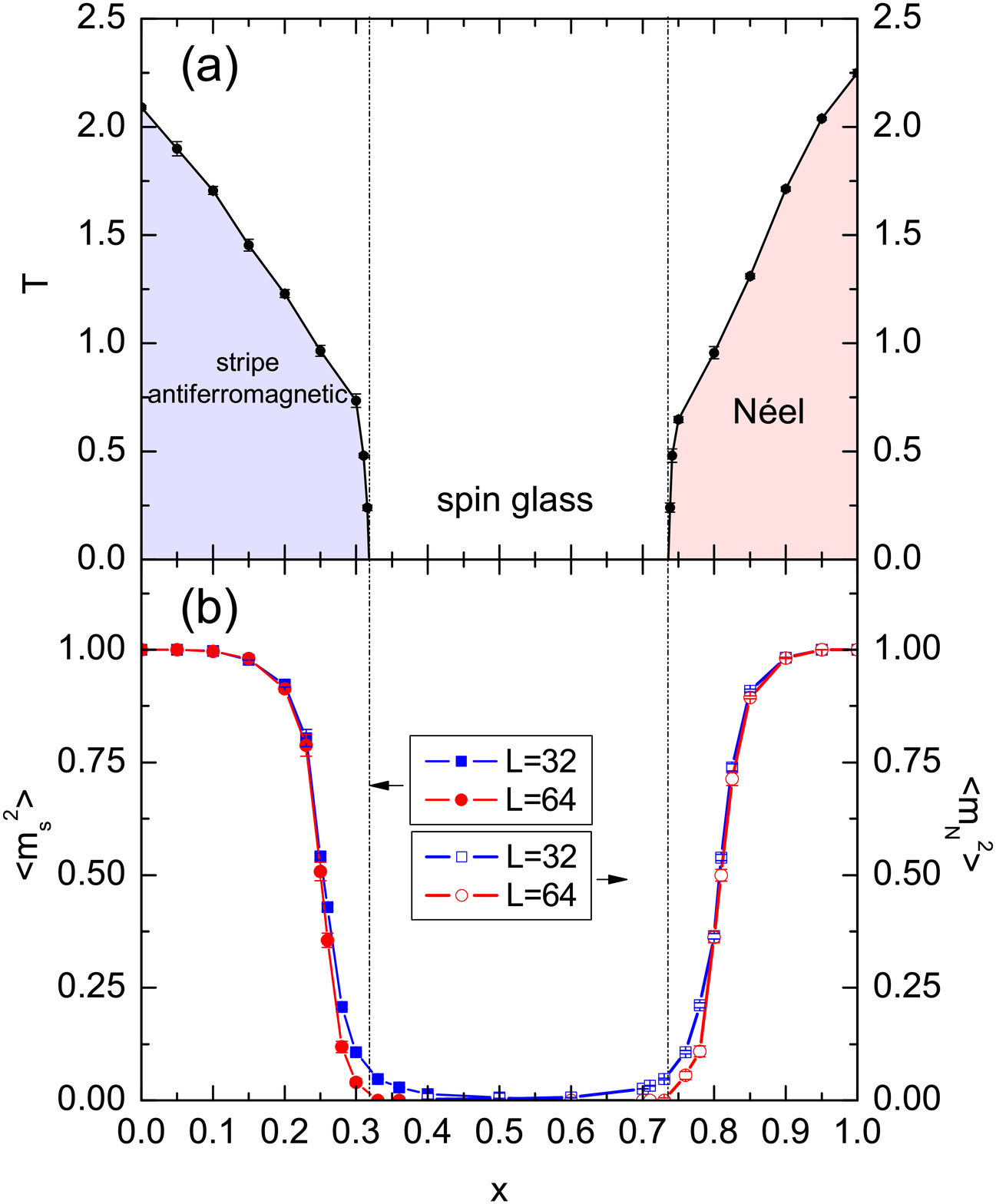}
\caption{(Color online) (a) is the phase diagram. By computing the critical temperature through a series dilution ratio $x$, we get two smooth phase boundaries. Below the critical temperatures, the system is in the stripe antiferromagnetic phase or the N\'{e}el phase, while the upper region is the paramagnetic phase. When the $x$ changes from the both sides to the middle, the critical temperature gradually decreases until the critical point can not be measured due to the disappearance of the long-range order. Since we can not find the $x_{c}$ value of $T=0$ by using classical MC, the $x_{c}$ at $T=0$ is obtained by extrapolating from the neighboring points. Here the $x_{c}$ of the stripe antiferromagnetic phase is $0.31(1)$, and the $x_{c}$ of the N\'{e}el phase is $0.73(2)$. (b) The results of order parameters $m_{s}^{2}$ and $m_{N}^{2}$. $T = 0.00003$ is used. The solid points show the results of $m_{s}^{2}$, and the open points correspond to $m_{N}^{2}$.}
\label{order3}
\end{figure}

When changing the dilution ratio $x$, we can obtain a series of critical points. From Fig.~\ref{order3}(a) we find that, as the disorder increasing, the critical temperature of the system decreases continuously to $0$ where the long-range order disappears. By sweeping the whole range of $x$ in $[0,1]$, we can find that the stripe antiferromagnetic phase has the $x_{c}=0.31(1)$, and the N\'{e}el phase has the $x_{c}=0.73(2)$. The phase diagram obtained here can help to understand the experimental phase diagram of $\mathrm{BaFe}_{2}(\mathrm{As}_{1-x}\mathrm{P}_{x})_{2}$, as discussed in Sec.~\ref{dscus}.

\subsection{The spin-glass phase \label{SG}}

In the following, we study the spin-glass phase which appears in the intermediate region of the $x$ ($0.31 < x < 0.73$).

The Edwards-Anderson (EA) order parameter $q$~\cite{Edwards1975SG} is defined as
  \begin{equation}
  q=\langle \sigma_{i}^{(1)} \sigma_{i}^{(2)}\rangle,
  \end{equation}
which measures the auto-correlation of spin $\sigma_{i}$ between the two replicas. As pointed out in Ref.\cite{Edwards1975SG}, the spin glass should have the characteristics: the magnetization $|m| = 0$ and the EA order parameter $|q| > 0$. The magnetization here is $m_{s}$ or $m_{N}$ which is used as the order parameter of the ordered phase, while the results of $m_{s}$ or $m_{N}$ are shown in Fig.~\ref{order3}(b). The results of EA order parameter are described in Sec~\ref{sp-e} where shows the existence of the spin-glass phase.

\subsubsection{Equilibrium finite-size scaling\label{sp-e}}

Here, we discuss the differences among the 2D ISG models. From the perspective of energy landscape, the 2D ISG model with Gaussian coupling has a non-degenerate ground state, implying the EA order parameter $|q|_{T = 0} = 1$, while the 2D $\pm J$ ISG model has infinitely degenerate ground states. From this point of view, our model is more similar to the 2D $\pm J$ ISG model, in which the $|q|$ can be expressed as a function of size when $T = 0$~\cite{XuNa2017Dual}.

\begin{figure}
\centering
\includegraphics[width=0.5\textwidth,height=2.2in]{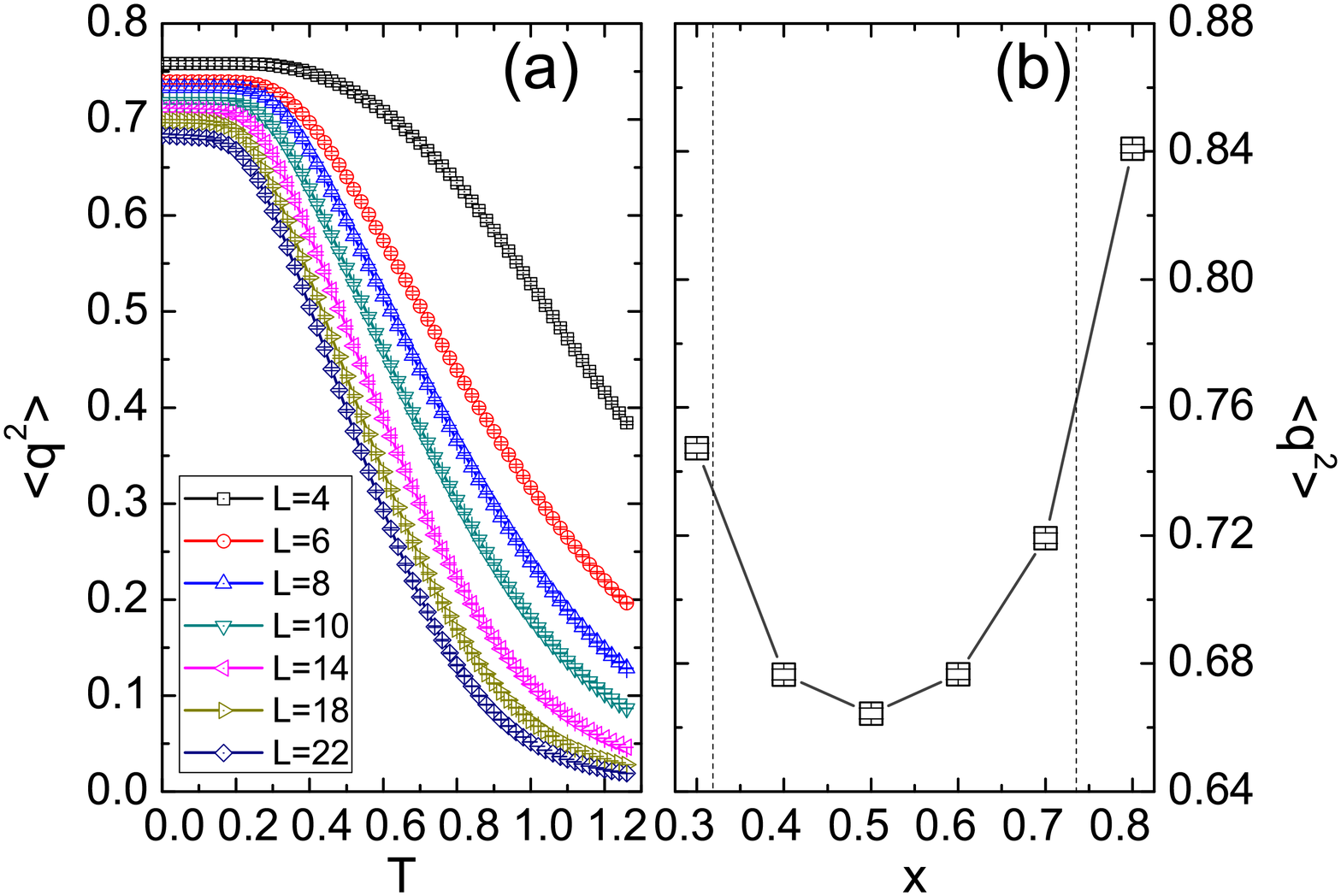}
\caption{(Color online) (a) shows the equilibrium $\langle q^{2} \rangle$ of the different system sizes vs $T$ when $x=0.5$. In the region of $T \rightarrow 0$, the $\langle q^{2} \rangle$ has a weak temperature dependence, and the values continue to decrease when the size increases. (b) The results of $\langle q^{2} \rangle$ vs $x$, while we use the dashed line to mark the spin-glass region. The minimum value appears at $x=0.5$. $L=24$ and $T=0.00002$ are used.}
\label{q-SG}
\end{figure}

As pointed out before, the 2D ISG only exists at $T = 0$, such that it is difficult for classical MC simulations to achieve the exact zero-temperature results. However one can still obtain the 2D ISG properties considering the weak dependence of the EA order parameter on temperature when $T \rightarrow 0$, as shown in Fig.\ref{q-SG}(a).

The EA order parameter is an important criterion to define and describe spin glass. Here we show the $\langle q^{2} \rangle$ values within the spin-glass region in Fig.\ref{q-SG}(b). Since the minimum of $\langle q^{2} \rangle$ appearing at $x=0.5$ where represents the most disordered state of the model, we take the fixed value $x=0.5$ as a sample in the following discussion.

\begin{figure}
\centering
\includegraphics[width=0.43\textwidth,height=3.5in]{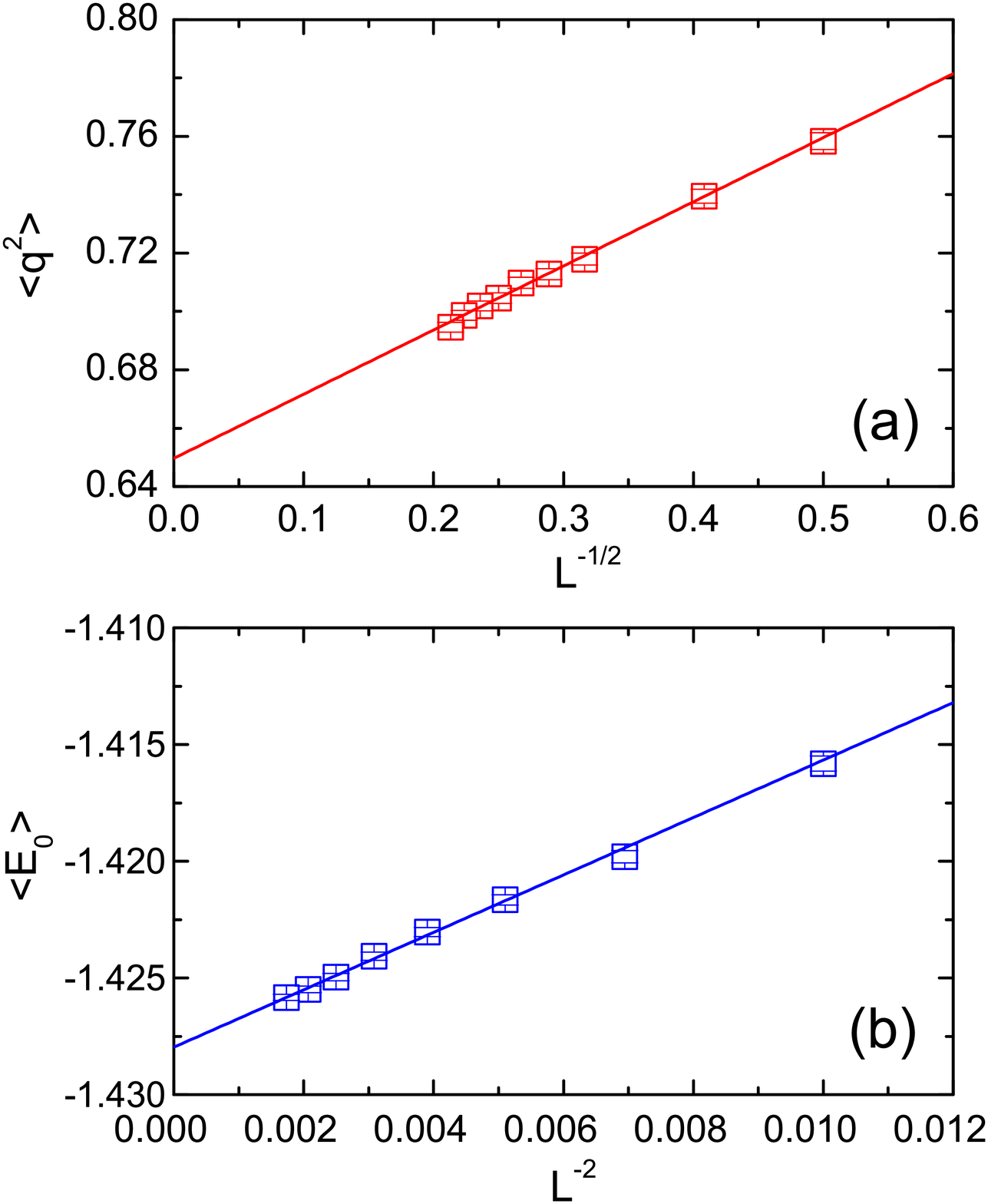}
\caption{(Color online). We show the equilibrium results $\langle q^{2}_{\mathrm{eq}}(L) \rangle$ (a) and $\langle E_{0}(L) \rangle$ (b) of $x = 0.5$ and $T \rightarrow 0$. Extrapolating the data to infinite size by the form of size correction, we obtain the results $\langle q^{2}_{\mathrm{eq}}(\infty) \rangle = 0.649(4)$ and $\langle E_{0}(\infty) \rangle = -1.4279(2)$.}
\label{rescale}
\end{figure}

We use the finite-size scaling relation~\cite{XuNa2017Dual} $A(T,L)=L^{-\kappa\Theta_{s}}f(TL^{\Theta_{s}})$ where $\Theta_{s}$ is the entropy exponent. The value of $\Theta_{s}$ in 2D $\pm J$ ISG model~\cite{G2011Thomas} is obtained by scaling spin-glass correlation function. We obtain the same result $\Theta_{s} \cong 0.5$ by performing the same scaling as Ref.\cite{G2011Thomas} in our model which shown in Appendix ~\ref{ap-A}. In order to find the value of $\langle q^{2}_{\mathrm{eq}} \rangle$ in the thermodynamic limit, we include a finite-size correction term~\cite{XuNa2017Dual}:
  \begin{equation}
  \langle q^{2}_{\mathrm{eq}}(L) \rangle - \langle q^{2}_{\mathrm{eq}}(\infty) \rangle \propto L^{-\Theta_{s}}.
  \label{scal-q}
  \end{equation}
In Ref.\cite{Campbell2004Energy}, the ground-state energy of spin glass is given by the finite-size correction, which is eventually written as $E(L)_{0} - E(\infty)_{0} \propto L^{-(d+ 1/\nu)}$. The $\nu$ of the case we studied here is discussed in Appendix ~\ref{ap-A}. Thus, we have $\nu \rightarrow \infty$, which leads to the correction of the ground-state energy to be written as
  \begin{equation}
  \langle E_{0}(L) \rangle - \langle E_{0}(\infty) \rangle \propto L^{-2} \qquad (d=2).
  \label{scal-E}
  \end{equation}
From the correction of the EA order parameter and the ground-state energy, we obtain the size-dependent relationship and the thermodynamic limit of $\langle q^{2}_{\mathrm{eq}}(L) \rangle$ and $\langle E_{0}(L) \rangle$ as shown in Fig.~\ref{rescale}. Here we obtain $\langle q^{2}_{\mathrm{eq}}(\infty) \rangle = 0.649(4)$ and $\langle E_{0}(\infty) \rangle = -1.4279(2)$. From the value of $\langle q^{2}_{\mathrm{eq}}(\infty) \rangle$ is greater than zero, according to the characteristic of the spin glass: $|m| = 0$ and $|q| > 0$, we can confirm that the spin-glass phase exists between the two ordered phases.

\subsubsection{Kibble-Zurek scaling\label{sp-a}}
In the calculations, a large number of updates are required to approach the equilibrium state, especially as $T \rightarrow 0$. Therefore, we consider the model at $x=0.5$ by using simulated annealing, which allows us anneal the system to $T = 0$ quickly and obtain the non-equilibrium results.

For nonlinear annealing\cite{Liu2014Dynamic}, we have
  \begin{equation}
  T=\upsilon(t_{\mathrm{max}}-t)^{r},
  \label{annealingtemperature}
  \end{equation}
where $\upsilon$ is the annealing velocity. The $\upsilon$ can be defined as $\upsilon=(T_{\mathrm{ini}}-T_{c})/t_{\mathrm{max}}^{r}$, where $t_{\mathrm{max}}$ is the number of the total MC annealing steps from an initial temperature $T_{\mathrm{ini}}$ to the critical temperature $T_{c}$, which is zero for the 2D ISG. The critical annealing velocity can be obtained from the Kibble-Zurek mechanism and expressed as $\upsilon_{\mathrm{KZ}}(L)\propto L^{-(zr+\Theta_{s})}$. When the annealing velocity $\upsilon$ is slower than the critical velocity $\upsilon_{\mathrm{KZ}}(L)$, the Kibble-Zurek scaling form of a singular quantity can be written by the annealing velocity and the system size:
  \begin{equation}
  A(\upsilon,L)=L^{-\kappa\Theta_{s}}F(\upsilon/\upsilon_{KZ})=L^{-\kappa\Theta_{s}}F(\upsilon L^{zr+\Theta_{s}}).
  \label{annealing}
  \end{equation}
The $z$ is the dynamic exponent which is defined by the relaxation time $\tau$ and the equilibrium-spatial-correlation length $\xi$:
  \begin{equation}
  \tau \propto \xi^{z}.
  \end{equation}

In order to ensure the correctness of the annealing results, we regenerate the distribution of $J_{2}$ before each annealing. To ensure that the annealing process starts from the paramagnetic state, we use a sufficiently high $T_{\mathrm{ini}} = 5 $. According to Eq (\ref{annealingtemperature}), we perform the annealing from the initial temperature to zero temperature. For the accuracy of results, numerous statistics are necessary, so we perform thousands of the annealing processes to average the results.

\begin{figure}
\centering
\includegraphics[width=0.45\textwidth,height=3.8in]{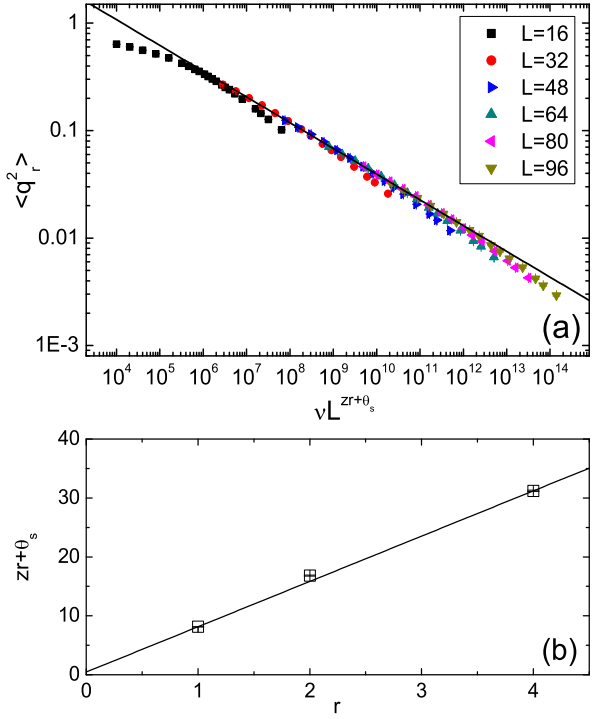}
\caption{(Color online) (a) Rescaling the $\langle q^{2} \rangle$ by $\langle q^{2} \rangle/\langle q^{2}_{\mathrm{eq}} \rangle$ and fitting the data, we can obtain the result of $zr+\Theta_{s}$ when $r=1$. (b) shows the results $zr+\Theta_{s}$ vs $r$. Annealing with a series of different $r$, we can obtain $z_{q} =7.68(4)$ and $\Theta_{s}=0.48(4)$.}
\label{annealing-q}
\end{figure}

Before fitting the results, we rewrite the Kibble-Zurek scaling form of the EA order parameter as
  \begin{equation}
  \langle q^{2}(\upsilon,L)\rangle = \langle q^{2}_{\mathrm{eq}}(L) \rangle F(\upsilon L^{zr+\Theta_{s}}),
  \label{q-annealing}
  \end{equation}
where $\langle q^{2}_{\mathrm{eq}}(L) \rangle$ is given by the previous equilibrium finite-size scaling form in Eq (\ref{scal-q}). When $\upsilon \rightarrow 0$, $\langle q^{2}(\upsilon,L) \rangle \rightarrow \langle q^{2}_{\mathrm{eq}}(L) \rangle$. In Fig.~\ref{annealing-q}(a), we define that $\langle q^{2}_{r} \rangle = \langle q^{2} \rangle/\langle q^{2}_{\mathrm{eq}} \rangle$, and the results can be well fitted on a straight line for $\upsilon < \upsilon_{\mathrm{KZ}}(L)$. By taking $r$ equals to 1, 2 and 4, we obtain the results of $z$ and $\Theta_{s}$ from $zr+\Theta_{s}$, as shown in Fig.~\ref{annealing-q}(b).

\begin{figure}
\centering
\includegraphics[width=0.45\textwidth,height=3.8in]{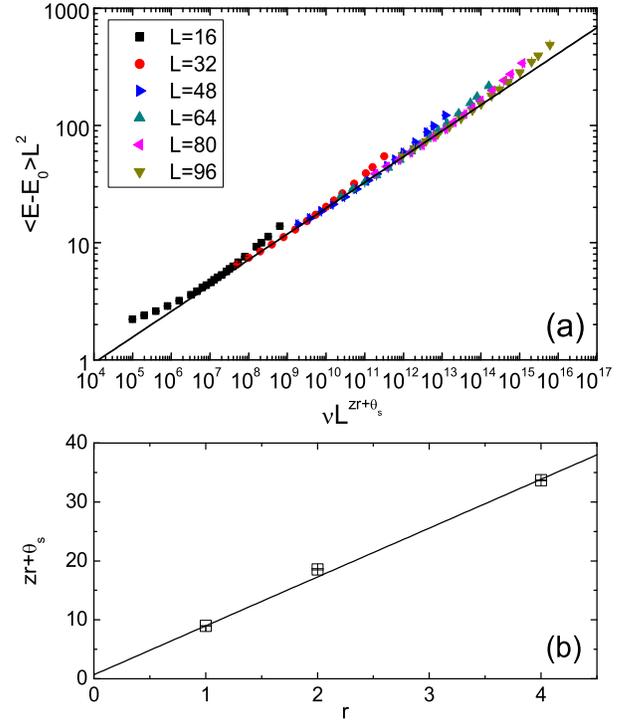}
\caption{(Color online) (a) We use the mean energy minus the equilibrium ground-state energy of infinite size and mulyiplied $L^{2}$ vs $\upsilon L^{zr+\Theta_{s}}$. Here are the results of $r=1$. (b) The results of $zr+\Theta_{s}$ vs $r$. We perform the same treatment as Fig.~\ref{annealing-q}(b), but the result is different: $z_{E} =8.44(2)$, $\Theta_{s}=0.51(3)$.}
\label{annealing-E}
\end{figure}

The Kibble-Zurek scaling form of the mean energy can be written as
  \begin{equation}
  \langle E(\upsilon,L) - E_{0}(\infty) \rangle = L^{-2} F(\upsilon L^{zr+\Theta_{s}}).
  \label{E-annealing}
  \end{equation}
The annealing results of the mean energy are shown in Fig.~\ref{annealing-E}, which are analyzed by the same technique as in Fig.~\ref{annealing-q}.

By checking the finite-size corrections to the scaling form of the EA order parameter and the mean energy, we obtain two different dynamic exponents $z_{q}$ and $z_{E}$ for the mean energy and the EA order parameter respectively. As Fig.~\ref{annealing-q} and Fig.~\ref{annealing-E} show, results of different annealing velocities and different sizes can be rescaled according to Eq (\ref{q-annealing}) and Eq (\ref{E-annealing}). The results $zr+\Theta_{s}$ of three different annealing paths $r = 1, 2, 4$ are consistent, and the entropy exponent obtained from $z_{q}r+\Theta_{s}$ and $z_{E}r+\Theta_{s}$ confirm $\Theta_{s} \cong 0.5$.

A similar situation was found in the 2D $\pm J$ ISG model shown in Ref.\cite{XuNa2017Dual}, where a detailed explanation is given by using the droplet theory. Here we discuss this unusual situation from the characteristic of the spin glass. The spin glass has short-range order and long-range disorder, implying the spin glass has some ordered clusters, while the clusters have no correlation among them. When the ordered clusters show up, the EA order parameter begins to access the stabilized value. Considering the energy landscape of spin glass, the clusters are ordered in both the ground state and the metastable state (local minimums). In the annealing process with the slow velocity, the system first enters the metastable state where the value of the EA order parameter begins to stabilize, but the energy continues to change until the system finally reaches the ground state. Therefore, the relaxation time $\tau$ for the EA order parameter is shorter than for the energy, while for the same system (with same correlation length $\xi$) the dynamic exponent $z_{q}$ is smaller than $z_{E}$.

\section{Discussion\label{dscus}}

The model we studied here has a very similar structure with the $\mathrm{FeAs}$ plane of the iron-based superconducting material $\mathrm{BaFe}_{2}(\mathrm{As}_{1-x}\mathrm{P}_{x})_{2}$, as detailed in Sec.~\ref{md}. The superconducting behaviors of $\mathrm{BaFe}_{2}(\mathrm{As}_{1-x}\mathrm{P}_{x})_{2}$ have already been studied in experiments \cite{BaFeAsP2009,MagneticBaFeAsPH2015,Unconventional2010Nakai}. Our research focuses on the magnetism, however it still has important implications to the real materials. Comparing the Fig.~\ref{order3}(a) with the experimental phase diagram for superconductivity of $\mathrm{BaFe}_{2}(\mathrm{As}_{1-x}\mathrm{P}_{x})_{2}$ in Ref.~\cite{BaFeAsP2009,Unconventional2010Nakai,MagneticBaFeAsPH2015}, we can get some interesting coincidence.

The stripe antiferromagnetic phase in Fig.~\ref{order3}(a) ($0\leq x < 0.31$) has a similar distribution with the phase obtained from the experiment which has antiferromagnetic order. Even the critical point of the stripe antiferromagnetic phase $x_{c}=0.31(1)$ is very similar to the experimental results. In the right side of the phase diagram, we obtain an N\'{e}el phase in $0.73 < x \leq 1$. It is possible that the N\'{e}el phase can suppress the appearance of the superconductivity. Our spin-glass phase appears at $0.31 < x < 0.73$, where exists superconductivity and the long-range magnetic order does not exist. The spin-glass phase is a special magnetic phase which does not exhibit the global magnetism and the long-range order. This character provides an advantageous environment for the emergence of superconductivity. The simultaneous appearance of superconductivity and spin glass was claimed in other superconducting materials~\cite{SGandSC2010}. For $\mathrm{BaFe}_{2}(\mathrm{As}_{1-x}\mathrm{P}_{x})_{2}$, the spin-glass-like behavior was suggested by the NMR and TRISP measurements for samples near the optimal region \cite{MagneticBaFeAsPH2015}.

\section{conclusions\label{conclu}}

By using the two different MC methods (parallel tempering and simulated annealing), we study a bond-diluted $J_{1}-J_{2}$ Ising model by changing the dilution ratio $x$ from 0 to 1. There are both the frustration and disorder in the model. By using the parallel temperature MC, different thermodynamic quantities are calculated for different $x$, and an interesting phase diagram is found in which a spin-glass phase exists. In the region $0\leq x < 0.31$, a stripe antiferromagnetic phase is found due to the frustration and the order on the system is not completely broken. In the region of $0.73 < x \leq 1$, the system maintains the N\'{e}el order as in $x = 1$ until the dilution ratio reaches the critical point. The spin-glass phase is found in the region $0.31 < x < 0.73$ and discussed from the equilibrium finite-size scaling where the scaling forms are similar with the 2D $\pm J$ ISG model. We perform the simulated annealing at the typical value $x = 0.5$, from that we obtain two different dynamic exponents $z_{q}$ and $z_{E}$ of the mean energy and the EA order parameter respectively by using the Kibble-Zurek mechanism. It is an unusual phenomenon, two different dynamic exponents are obtained from the same annealing process. Since we have obtained some results similar to the 2D $\pm J$ ISG model, which allows us to connect our model with the the 2D $\pm J$ ISG model which is a classical 2D ISG model, even though the distribution of interactions is very different.

The magnetism of iron-based superconducting material $\mathrm{BaFe}_{2}(\mathrm{As}_{1-x}\mathrm{P}_{x})_{2}$ on $\mathrm{Fe}$ can be described by the model we studied here. In this material, the next-nearest-neighbor interactions $J_{2}$ are controlled by the $\mathrm{As}$ atoms, which can be substituted with the $\mathrm{P}$ atoms. An interesting discovery is that our phase diagram is similar to the experimental phase diagram of $\mathrm{BaFe}_{2}(\mathrm{As}_{1-x}\mathrm{P}_{x})_{2}$, which helps us to understand the magnetism behind the material.

\begin{acknowledgments}
We thank A. W. Sandvik, E. W. Carlson, W. F. Tsai, E. Dagotto, J. P. Hu, and N. Xu for helpful discussions. This project is supported by NKRDPC-2017YFA0206203, NSFC-11574404, NSFC-11275279, NSFG-2015A030313176, Special Program for Applied Research on Super Computation of the NSFC-Guangdong Joint Fund, National Supercomputer Center In Guangzhou, and Leading Talent Program of Guangdong Special Projects.
\end{acknowledgments}

\appendix
\section{Connections with the 2D $\pm J$ ISG model\label{ap-A}}

The 2D $\pm J$ ISG model is a classical spin-glass model. As mentioned in Sec.~\ref{sp-e}, the properties of the 2D ISG are different in different models. Since the spin glass here has a similar behavior as the 2D $\pm J$ ISG model with the EA order parameter, so we pay more attention to the scaling form of 2D $\pm J$ ISG model for other physical quantities.

Thomas et al~\cite{droplet1988Fisher} used the droplet theory to discuss the 2D $\pm J$ ISG model in Ref.~\cite{G2011Thomas}. They gave a scaling of the correlation function $G(\vec{r})$ at $T\rightarrow0$ by using the entropy exponent $\Theta_{s}$. The correlation function $G_{0}(\vec{r})=[\langle \sigma_{\vec{0}}\sigma_{\vec{r}} \rangle^{2}_{0}]$ at large $r$ behaved as
  \begin{equation}
  G_{0}(\vec{r})-G_{0}(\infty) \thicksim r^{-\Theta_{s}},
  \end{equation}
where $\Theta_{s} \cong 0.5$ and $r$ represents the distance between two spins.

\begin{figure}
\centering
\includegraphics[width=0.5\textwidth,height=2.4in]{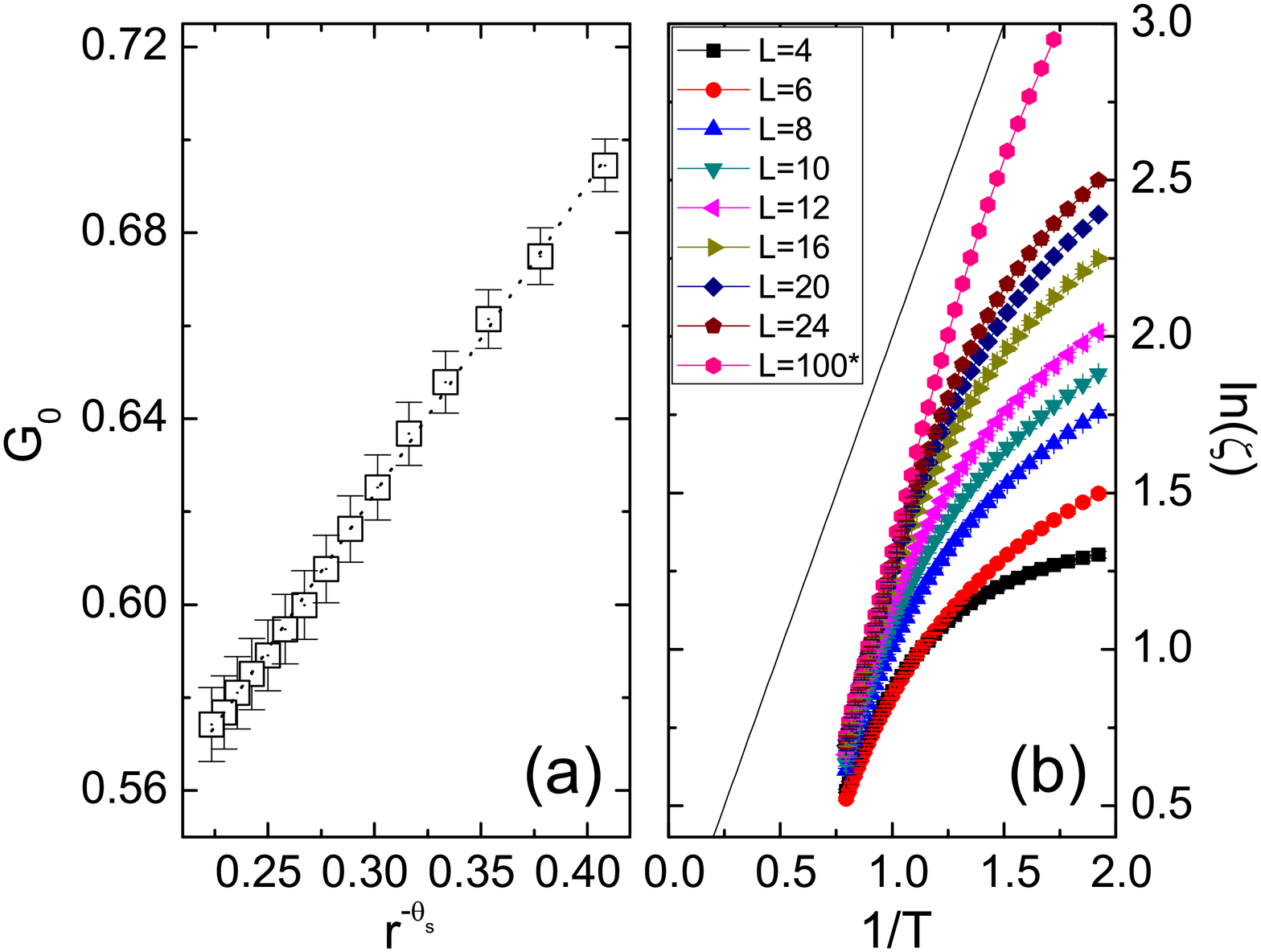}
\caption{(Color online). (a) The behavior of the correlation function $G_{0}$ at large $r$. Here $x=0.5$, $L=48$, $T=0.00002$. By using $\Theta_{s}=0.5 $, we can get a good linear fitting, so we can say that here $\Theta_{s} \cong 0.5$ is also established. (b)$\ln(\xi)$ vs $1/T$, which reflects the correlation length diverging exponentially at $x=0.5$. We can extrapolate the results for $L=100$ and deduce that the $\xi$ has the form of $\xi \sim \exp(2 \beta J)$ when $T \rightarrow 0$ and $L \rightarrow \infty$. Finally, we can get $\nu \rightarrow \infty$.
\label{G-xi}}
\end{figure}

To determine the value of $\Theta_{s}$ in the spin glass here, we perform the same scaling for $G_{0}(\vec{r})$, as shown in Fig.~\ref{G-xi}(a). From the scaling results, we find $\Theta_{s} \cong 0.5$ is applicable here. So in Eq (\ref{scal-q}), we use $0.5$ as the value of $\Theta_{s}$ when we perform the scaling of $\langle q^{2} \rangle$.

Correlation length is a commonly used physical quantity in the study of spin glass. The value of the critical exponent $\nu$ can be determined by $\xi \sim |T - T_{c}|^{-\nu}$. In the MC simulations, $\xi$ can be obtained from the susceptibility of spin glass $\chi_{\mathrm{SG}}$:
\begin{eqnarray}
  \chi_{\mathrm{SG}}(\mathbf{k})=\frac{1}{N} \sum\limits_{i,j} [\langle \sigma_{i} \sigma_{j} \rangle ^{2}]_{\mathrm{av}} e^{i\mathbf{k}\cdot(\mathbf{R}_{i}-\mathbf{R}_{j})},\\
  \xi_{L}=\frac{1}{2\sin(|\mathbf{k}_{\mathrm{min}}|/2)}\left[\frac{\chi_{\mathrm{SG}}(0)}{\chi_{\mathrm{SG}}(\mathbf{k}_{\mathrm{min}})}-1\right]^{1/2},
\end{eqnarray}
where $\mathbf{k}_{\mathrm{min}}=(2\pi/L,0)$.

In the study of the 2D $\pm J$ ISG model, one of the points is $\nu \rightarrow \infty$, like in Ref.~\cite{Correlation2005Katzgraber}. We confirm that the correlation length $\xi$ in our study is exponentially diverged by performing the same analysis as in the Ref.~\cite{Correlation2005Katzgraber}, and the results are shown in Fig.~\ref{G-xi}(b). Therefore, we also have the result of $\nu \rightarrow \infty$ here. This result is used for the equilibrium finite-size scaling form of ground-state energy in the Eq (\ref{scal-E}).

\bibliographystyle{apsrev4-1}
\newpage
\bibliography{reference}
\end{document}